\documentclass[preprint2]{aastex}

\usepackage{graphicx}
\usepackage{epstopdf}
\usepackage{color}
\usepackage{natbib}
\usepackage{hyperref} 
\usepackage{longtable} 
\usepackage{epsf} 
\usepackage{breakurl}
\usepackage{breakcites}
\usepackage{afterpage}

\shorttitle{Search and Characterization of Wide Binary Systems}
\shortauthors{Baron et al.}

\begin{document}


\title{DISCOVERY AND CHARACTERIZATION OF WIDE BINARY SYSTEMS WITH A VERY LOW MASS COMPONENT}


\author{Fr\'ed\'erique Baron\footnote{baron@astro.umontreal.ca}\altaffilmark{1}, David Lafreni\`ere\altaffilmark{1}, \'Etienne Artigau\altaffilmark{1}, Ren\'e Doyon\altaffilmark{1},  Jonathan Gagn\'e\altaffilmark{1}, Cassy L. Davison\altaffilmark{2}, Lison Malo\altaffilmark{3}, Jasmin Robert\altaffilmark{1}, Daniel Nadeau\altaffilmark{1}, and C\'eline Reyl\'e\altaffilmark{4} }
\affil{$^1$ D\'epartement de Physique, Universit\'e de Montr\'eal, C.P. 6128 Succ. Centre-ville, Montr\'eal, Qc H3C 3J7, Canada}
\affil{$^2$ Department of Physics and Astronomy, Georgia State University, Atlanta, GA 30303, USA}
\affil{$^3$ Canada-France-Hawaii Telescope, 65-1238 Mamalahoa Hwy, Kamuela, HI 96743, USA}
\affil{$^4$ Institut Utinam, CNRS UMR6213, Universit\'e de Franche-Comt\'e, OSU THETA Franche-Comt\'e-Bourgogne, Observatoire de Besan\c[c]on, BP 1615, 25010 Besançon Cedex, France}


\begin{abstract}
We report the discovery of 14 low-mass binary systems containing mid-M to mid-L dwarf companions with separations larger than 250~AU. We also report the independent discovery of 9 other systems with similar characteristics that were recently discovered in other studies. We have identified these systems by searching for common proper motion sources in the vicinity of known high proper motion stars, based on a cross-correlation of wide area near-infrared surveys (2MASS, SDSS, and SIMP). An astrometric follow-up, for common proper motion confirmation, was made with SIMON and/or CPAPIR at the OMM 1.6 m and CTIO 1.5 m telescopes for all the candidates identified. A spectroscopic follow-up was also made with GMOS or GNIRS at Gemini to determine the spectral types of 11 of our newly identified companions and 10 of our primaries. Statistical arguments are provided to show that all of the systems we report here are very likely to be physical binaries. One of the new systems reported features a brown dwarf companion: LSPM J1259+1001 (M5) has an L4.5 (2M1259+1001) companion at $\sim$340~AU. This brown dwarf was previously unknown. Seven other systems have a companion of spectral type L0--L1 at a separation in the 250--7500~AU range. Our sample includes 14 systems with a mass ratio below 0.3. 

\end{abstract}


\keywords{binaries: general --- stars: low-mass, brown dwarfs }

\section{INTRODUCTION}

In the last two decades, the search for low-mass stars and brown dwarfs has intensified with the advent of wide area surveys such as 2MASS \citep{cutri_2mass_2003}, DENIS \citep{epchtein_preliminary_1999}, SDSS \citep{ahn_ninth_2012}, UKIDSS \citep{lawrence_ukirt_2007}, and more recently WISE \citep{cutri_wise_2012}. In addition, a handful of surveys were aimed at finding nearby binary systems including low mass stars such as the surveys of \cite{reid_low-mass_1997} and \cite{reid_search_2001} for example.  \cite{dhital_sloan_2010} assembled the SLoWPoKES catalog which contains 1342 binary systems with a projected separation of $\gtrsim$ 500 AU, in which one of the components is a mid-K to mid-M dwarf. Numerous wide binary systems have also been discovered individually. Of all the systems known, very few have a separation larger than 7000 AU.  The system composed of NLTT 20346 (an M5V and M6V binary) and 2MASS J0850359+105716 (an L6 dwarf) is one of these, with a separation of 7700 AU \citep{radigan_discovery_2009}, while Konigstuhl 3 A-BC \citep{faherty_identification_2011}, a hierarchical system composed of an F8V orbited by an M8V and L3 binary, has a separation of 12 000 AU. 

An interest of binary systems is that both components share the same age and metallicity. For systems comprising an M dwarf companion for example, this can help calibrate the metallicity scale for this type of star \citep{bonfils_metallicity_2005}. The components of wide visual binaries can also be studied in detail individually using seeing-limited instruments, making their characterization much easier. Additionally, the evolution of each component of a wide binary system has not been influenced by the other component, and both can be viewed as an ``isolated" star.

The discovery of numerous low-mass stars and brown dwarfs has led to a growing interest in understanding the formation mechanism of these objects whose masses are below the typical Jeans mass in a molecular cloud \citep{meyer_stellar_2000}, and numerous hypotheses about their origin have been formulated over the years. Binary systems containing a low mass component provide a good tool for attempting to discriminate between the various possibilities, as their properties (e.g., separations, mass ratios) may be affected differently by different processes \citep{duchene_stellar_2013}. Wide binary systems with separation larger than hundreds of AU are of particular interest for this purpose as, being more weakly gravitationally bound and spanning sizes larger than typical protostellar disks, they bring additional and perhaps more stringent constraints on the underlying formation processes.

In this article, we report the discovery of 14 new binary systems that all have a component with a spectral type later than M6 and a separation larger than 250~AU. We also recovered nine previously known binaries  with similar characteristics. Section \ref{sec:discovery} details our archival search for wide binaries and the identification of our candidates, while Section \ref{sec:imaging} describes the follow-up photometry and the calculations of the proper motion. The  spectroscopic follow-up as well as the results extracted from these data are explained in Section \ref{sec:spectro}. The probability of random alignement has been calculated for each system and is presented in Section \ref{sec:prob}. A short discussion is presented in section \ref{sec:discussion}.  Our results are summarized in section \ref{sec:conclu}.

\begin{deluxetable}{llcccccccc}  

\tabletypesize{\tiny}
\tablewidth{0pt}
\rotate
\tablecolumns{10}
\tablecaption{Photometric data\label{photo}}
\tablehead{   
  \colhead{Name} &
  \colhead{2MASS Name} &
  \colhead{$\alpha^{\textrm{a}}$} &
  \colhead{$\delta^{\textrm{a}}$}&
  \colhead{$r'^{\textrm{b}}$} &
  \colhead{$i'^{\textrm{b}}$}&
  \colhead{$z'^{\textrm{b}}$} &
  \colhead{$J^{\textrm{a}}$}&
  \colhead{$H^{\textrm{a}}$} &
  \colhead{$K_s^{\textrm{a}}$}

}
\startdata
NLTT 251$^{\textrm{c}}$ &2MASS J00064746-0852350&$00^h06^m47^s.47$&$-08\,^{\circ}$ $ 52^{\prime}$ $35.1^{\prime\prime}$&$17.30\pm0.01$&$14.94\pm0.01$&$13.73\pm0.01$&$11.97\pm0.02$&$11.43\pm0.02$&$11.09\pm0.02$\\
\nodata &2MASS J00064916-0852457 &$00^h06^m49^s.17$&$-08\,^{\circ}$ $52^{\prime}$ $45.7^{\prime\prime}$&$20.91\pm0.05$&$18.26\pm0.01$&$16.45\pm0.01$&$14.14\pm0.03$&$13.55\pm0.03$&$13.13\pm0.04$\\\hline

NLTT 687&2MASS J00140212-1814578&$00^h14^m02^s.10$&$-18\,^{\circ}$ $14^{\prime}$ $58.0^{\prime\prime}$&\nodata &\nodata &\nodata &$10.38\pm0.02$&$9.81\pm0.02$&$9.57\pm0.02$\\
\nodata &2MASS J00135882-1816462&$00^h13^m58^s.80$&$-18\,^{\circ}$ $16^{\prime}$ $46.2^{\prime\prime}$&\nodata &\nodata &\nodata &$16.54\pm0.14$&$15.89\pm0.18$&$15.04\pm0.13$\\\hline

NLTT 2274$^{\textrm{d}}$&2MASS J00415543+1341162&$00^h41^m55^s.43$&$+13\,^{\circ}$ $41^{\prime}$  $16.2^{\prime\prime}$ &\nodata $^{\textrm{g}}$&$15.42\pm0.01$&\nodata $^{\textrm{g}}$&$10.16\pm0.03$&$ 9.57\pm0.03$&$ 9.35\pm0.02$\\
\nodata &2MASS J00415453+1341351&$00^h41^m54^s.53$&$+13\,^{\circ}$ $41^{\prime}$ $35.1^{\prime\prime}$ &$21.15\pm0.07$&$18.71\pm0.01$&$16.83\pm0.01$&$14.45\pm0.03$&$13.67\pm0.04$&$13.24\pm0.03$\\\hline

BD-06 813 &2MASS J04050271-0600261 &$04^h05^m02^s.71$&$-06\,^{\circ}$ $00^{\prime}$ $26.1^{\prime\prime}$&\nodata $^{\textrm{g}}$&\nodata $^{\textrm{g}}$&\nodata $^{\textrm{g}}$&$8.35\pm0.02$&$7.93\pm0.03$&$7.87\pm0.03$\\
\nodata &2MASS J04050209-0600409&$04^h05^m02^s.10$&$-06\,^{\circ}$ $00^{\prime}$ $40.9^{\prime\prime}$&$21.22\pm0.06$&$18.54\pm0.01$&$17.08\pm0.01$&$15.11\pm0.05$&$14.30\pm0.05$&$13.88\pm0.05$\\\hline

NLTT 20640$^{\textrm{e}}$&2MASS J08583671+2711068 &$08^h58^m36^s.72$&$+27\,^{\circ}$ $11^{\prime}$ $06.8^{\prime\prime}$ &$15.92\pm0.01$&\nodata $^{\textrm{g}}$&$13.50\pm0.01$&$11.99\pm0.02$&$11.40\pm0.02$&$11.10\pm0.01$\\
\nodata &2MASS J08583693+2710518&$08^h58^m36^s.94$&$+27\,^{\circ}$ $10^{\prime}$ $51.8^{\prime\prime}$&$22.05\pm0.11$&$19.45\pm0.02$&$17.64\pm0.02$&$15.05\pm0.05$&$14.23\pm0.05$&$13.66\pm0.05$\\\hline

LSPM J1021+3704&2MASS J10215240+3704289&$10^h21^m52^s.40$&$+37\,^{\circ}$ $04^{\prime}$ $29.0^{\prime\prime}$ &$16.07\pm0.01$&\nodata $^{\textrm{g}}$&$13.87\pm0.01$&$12.43\pm0.02$&$11.95\pm0.02$&$11.76\pm0.02$\\
\nodata &2MASS J10215386+3704166&$10^h21^m53^s.87$&$+37\,^{\circ}$ $04^{\prime}$ $16.7^{\prime\prime}$ &$24.67\pm0.48$&$21.85\pm0.14$&$19.91\pm0.08$&$17.18\pm0.03^{\textrm{f}}$&$16.25\pm0.22$&$15.66\pm0.23$\\\hline

\nodata &2MASS J10432398-1706024&$10^h43^m23^s.99$&$-17\,^{\circ}$ $06^{\prime}$ $02.4^{\prime\prime}$&\nodata &\nodata &\nodata &$11.48\pm0.02$&$10.89\pm0.02$&$10.60\pm0.03$\\
\nodata &2MASS J10432513-1706065&$10^h43^m25^s.13$&$-17\,^{\circ}$ $06^{\prime}$ $06.6^{\prime\prime}$ &\nodata &\nodata &\nodata &$15.67\pm0.06$&$15.07\pm0.06$&$14.61\pm0.11$\\\hline

NLTT 26746&2MASS J11150134+1606447&$11^h15^m01^s.35$&$+16\,^{\circ}$ $06^{\prime}$ $44.7^{\prime\prime}$ &\nodata $^{\textrm{g}}$&$16.61\pm0.01$&\nodata $^{\textrm{g}}$&$11.14\pm0.02$&$10.56\pm0.02$&$10.32\pm0.02$\\
\nodata &2MASS J11150150+1607026&$11^h15^m01^s.50$&$+16\,^{\circ}$:$07^{\prime}$ $02.7^{\prime\prime}$&$22.94\pm0.37$&$20.71\pm0.07$&$18.59\pm0.05$&$16.40\pm0.12$&$15.22\pm0.09$&$14.56\pm0.11$\\\hline

NLTT 29392&2MASS J12024963+4204475&$12^h02^m49^s.63$&$+42\,^{\circ}$ $04^{\prime}$ $47.5^{\prime\prime}$ &$18.21\pm0.01$&\nodata $^{\textrm{g}}$&$13.87\pm0.01$&$12.43\pm0.02$&$11.95\pm0.02$&$11.67\pm0.02$\\
\nodata &2MASS J12025009+4204531&$12^h02^m50^s.09$&$+42\,^{\circ}$ $04^{\prime}$ $53.2^{\prime\prime}$&$21.45\pm0.07$&$18.87\pm0.01$&$16.96\pm0.01$&$14.50\pm0.03$&$13.69\pm0.03$&$13.26\pm0.03$\\\hline

LSPM J1259+1001&2MASS J12594217+1001407&$12^h59^m42^s.17$&$+10\,^{\circ}$ $01^{\prime}$ $40.7^{\prime\prime}$&$17.76\pm0.01$&$15.63\pm0.00$&$14.44\pm0.00$&$12.60\pm0.02$&$12.03\pm0.03$&$11.63\pm0.02$\\
\nodata &2MASS J12594167+1001380&$12^h59^m41^s.68$&$+10\,^{\circ}$ $01^{\prime}$ $38.06^{\prime\prime}$&$24.45\pm0.76$&$21.58\pm0.13$&$19.60\pm0.10$&$16.72\pm0.19$&$15.53\pm0.14$&$14.89\pm0.15$\\\hline

NLTT 36369&2MASS J14081918+3708294&$14^h08^m19^s.19$&$+37\,^{\circ}$ $08^{\prime}$ $29.4^{\prime\prime}$
&$15.60\pm0.01$&$14.53\pm0.01$&$13.48\pm0.01$&$12.06\pm0.02$&$11.45\pm0.02$&$11.22\pm0.02$\\
\nodata &2MASS J14081969+3708255&$14^h08^m19^s.70$&$+37\,^{\circ}$ $08^{\prime}$ $25.6^{\prime\prime}$&$21.98\pm0.10$&$19.09\pm0.02$&$17.47\pm0.02$&$15.42\pm0.05$&$14.78\pm0.06$&$14.35\pm0.08$\\\hline

LSPM J1441+1856&2MASS J14412209+1856451&$14^h41^m22^s.09$&$+18\,^{\circ}$ $56^{\prime}$ $45.1^{\prime\prime}$&$17.43\pm0.04$&\nodata $^{\textrm{g}}$&\nodata $^{\textrm{g}}$&$11.44\pm0.02$&$10.82\pm0.03$&$10.55\pm0.02$\\
\nodata &2MASS J14412565+1856484&$14^h41^m25^s.65$&$+18\,^{\circ}$ $56^{\prime}$ $48.5^{\prime\prime}$&$23.38\pm0.27$&$21.34\pm0.07$&$19.41\pm0.06$&$16.94\pm0.18$&$16.17\pm0.17$&$15.40\pm0.13$\\\hline

NLTT 41701&2MASS J15590815+3735477&$15^h59^m08^s.15$&$+37\,^{\circ}$ $35^{\prime}$ $47.7^{\prime\prime}$ &\nodata $^{\textrm{g}}$&\nodata $^{\textrm{g}}$&\nodata $^{\textrm{g}}$&$~9.12\pm0.02$&$~8.67\pm0.02$&$~8.56\pm0.02$\\
\nodata &2MASS J15590740+3735275&$15^h59^m07^s.40$&$+37\,^{\circ}$ $35^{\prime}$ $27.5^{\prime\prime}$&$22.89\pm0.27$&$19.66\pm0.03$&$18.07\pm0.03$&$15.75\pm0.07$&$14.92\pm0.07$&$14.53\pm0.08$\\\hline

HD 234344&2MASS J16460765+5020405&$16^h46^m07^s.64$&$+50\,^{\circ}$ $20^{\prime}$ $41.1^{\prime\prime}$ &\nodata $^{\textrm{g}}$&\nodata $^{\textrm{g}}$&\nodata $^{\textrm{g}}$&$7.62\pm0.02$&$7.06\pm0.02$&$6.97\pm0.03$\\
\nodata &2MASS J16461148+5019456&$16^h46^m11^s.48$&$+50\,^{\circ}$ $19^{\prime}$ $45.7^{\prime\prime}$ &$20.47\pm0.03$&$17.43\pm0.01$&$16.66\pm0.01$&$13.61\pm0.02$&$12.98\pm0.03$&$12.59\pm0.02$\\\hline

HD 217246&2MASS J22591790+0806484&$22^h59^m17^s.90$&$+08\,^{\circ}$ $06^{\prime}$ $48.4^{\prime\prime}$ &\nodata $^{\textrm{g}}$&\nodata $^{\textrm{g}}$&\nodata $^{\textrm{g}}$&$8.56\pm0.02$&$8.26\pm0.03$&$8.19\pm0.02$\\
\nodata &2MASS J22591631+0806556&$22^h59^m16^s.31$&$+08\,^{\circ}$ $06^{\prime}$ $ 55.6^{\prime\prime}$&$21.83\pm0.09$&$18.84\pm0.01$&$17.61\pm0.02$&$15.76\pm0.06$&$15.14\pm0.08$&$14.74\pm0.10$\\\hline

NLTT 56936 &2MASS J23274840+0451241 &$23^h27^m48^s.40$&$+04\,^{\circ}$ $51^{\prime}$ $24.2^{\prime\prime}$ &\nodata $^{\textrm{g}}$ &\nodata $^{\textrm{g}}$ &\nodata $^{\textrm{g}}$ &$8.16\pm0.03$&$7.58\pm0.05$&$7.41\pm0.02$\\
\nodata &2MASS J23274947+0450583&$23^h27^m49^s.48$&$+04\,^{\circ}$ $50^{\prime}$ $58.4^{\prime\prime}$&$21.54\pm0.125$ &$18.92\pm0.02$ &$17.17\pm0.02$ &$15.10\pm0.03$&$14.36\pm0.04$&$13.97\pm0.05$\\\hline

TYC 1725-344-1&2MASS J23553113+1755239&$23^h55m31^s.14$&$+17\,^{\circ}$ $55^{\prime}$ $23.9^{\prime\prime}$ &\nodata $^{\textrm{g}}$&\nodata $^{\textrm{g}}$&\nodata $^{\textrm{g}}$&$8.80\pm0.03$&$8.38\pm0.02$&$8.28\pm0.02$\\
\nodata &2MASS J2355345+175404 &$23^h55^m34^s.50$&$+17\,^{\circ}$ $54^{\prime}$ $04.0^{\prime\prime}$ &$23.24\pm0.33$&$20.43\pm0.04$&$18.73\pm0.03$&$16.39\pm0.12$&$15.42\pm0.15$&$15.02\pm0.15$\\\hline\hline

NLTT 182 &2MASS J00054143+0626256&$00^h05^m41^s.43$&$+06\,^{\circ}$ $26^{\prime}$ $25.6^{\prime\prime}$ &$17.08\pm0.01$&$15.42\pm0.04$&$14.51\pm0.01$&$13.02\pm0.03$&$12.51\pm0.03$&$12.19\pm0.03$\\
\nodata &2MASS J00054171+0626300&$00^h05^m41^s.71$&$+06\,^{\circ}$ $26^{\prime}$ $30.1^{\prime\prime}$ &$22.38\pm0.16$&$19.86\pm0.03$&$18.00\pm0.02$&$15.81\pm0.10$&$15.13\pm0.11$&$14.62\pm0.11$\\\hline

HD 2292&2MASS J00265848+1705088&$00^h26^m58^s.49$&$+17\,^{\circ}$ $05^{\prime}$ $08.8^{\prime\prime}$&$8.88\pm0.01$ &\nodata $^{\textrm{g}}$ &\nodata $^{\textrm{g}}$ &$7.90\pm0.027$&$7.55\pm0.04$&$7.50\pm0.03$\\
\nodata &2MASS J00265989+1704463&$00^h26^m59^s.90$&$+17\,^{\circ}$ $04^{\prime}$ $46.4^{\prime\prime}$&$21.82\pm0.13$ &$19.16\pm0.02$ &$17.39\pm0.02$ &$15.40\pm0.05$&$14.67\pm0.05$&$14.39\pm0.07$\\\hline

NLTT 4558 &2MASS J01221872+0330470 &$01^h22^m18^s.72$&$+03\,^{\circ}$ $30^{\prime}$:$47.1^{\prime\prime}$&\nodata $^{\textrm{g}}$&\nodata $^{\textrm{g}}$&\nodata $^{\textrm{g}}$&$7.42\pm0.02$&$7.14\pm0.03$&$7.07\pm0.02$\\
\nodata &2MASS J01221697+0331235&$01^h22^m16^s.98$&$+03\,^{\circ}$ $31^{\prime}$ $23.6^{\prime\prime}$&$22.05\pm0.13$&$19.95\pm0.03$&$17.95\pm0.03$&$15.47\pm0.05$&$14.65\pm0.06$&$14.40\pm0.08$\\\hline

NLTT 30510&2MASS J12221994+3643539&$12^h22^m19^s.94$&$+36\,^{\circ}$ $43^{\prime}$ $54.0^{\prime\prime}$ &\nodata $^{\textrm{g}}$&\nodata $^{\textrm{g}}$&\nodata $^{\textrm{g}}$&$10.51\pm0.02$&$9.98\pm0.02$&$9.73\pm0.02$\\
\nodata &2MASS J12221837+3643485&$12^h22^m18^s.37$&$+36\,^{\circ}$ $43^{\prime}$ $48.5^{\prime\prime}$ &$22.61\pm0.17$&$20.21\pm0.03$&$18.15\pm0.03$&$15.97\pm0.08$&$15.27\pm0.09$&$14.85\pm0.09$\\\hline

LSPM J1236+3000&2MASS J12363558+3000315&$12^h36^m35^s.59$&$+30\,^{\circ}$ $00^{\prime}$ $31.6^{\prime\prime}$ &$19.63\pm0.02$&$17.75\pm0.01$&$16.81\pm0.01$&$15.28\pm0.05$&$14.60\pm0.06$&$14.30\pm0.08$\\
\nodata &2MASS J12363647+3000315&$12^h36^m36^s.48$&$+30\,^{\circ}$ $00^{\prime}$ $31.5^{\prime\prime}$&$23.04\pm0.31$&$20.63\pm0.05$&$18.86\pm0.05$&$16.82\pm0.20$&$16.06\pm0.22$&$15.48\pm0.20$\\\hline

NLTT 33793$^{\textrm{d}}$ &2MASS J13205010+0955582&$13^h20^m50^s.11$&$+09\,^{\circ}$ $55^{\prime}$ $58.3^{\prime\prime}$ &\nodata $^{\textrm{g}}$&\nodata $^{\textrm{g}}$&\nodata $^{\textrm{g}}$&$7.89\pm0.03$&$7.32\pm0.04$&$7.22\pm0.03$\\
\nodata &2MASS J13204159+0957506&$13^h20^m41^s.59$&$+09\,^{\circ}$ $57^{\prime}$:$50.6^{\prime\prime}$&$20.40\pm0.03$&$17.47\pm0.01$&$15.80\pm0.01$&$13.73\pm0.03$&$13.08\pm0.03$&$12.61\pm0.03$\\\hline

\enddata

\tablenotetext{a}{Data are from the 2MASS catalog \citep{cutri_2mass_2003}}
\tablenotetext{b}{Data are from the SDSS catalog \citep{ahn_ninth_2012}}
\tablenotetext{c}{This system has been identifified to be a hierarchical triple system with an M6 primary and an M9+T5 secondary by \cite{burgasser_discovery_2012}}
\tablenotetext{d}{This system has been identifified as a binary by \citet{faherty_brown_2010}}
\tablenotetext{e}{This system has been identifified as a binary by \cite{zhang_discovery_2010}}
\tablenotetext{f}{This magnitude has been extrated from a CPAPIR MKO J-band image and has been converted to the 2MASS magnitude system. This object was poorly detected in the 2MASS calatog as it is faint.}
\tablenotetext{g}{Saturated}

\end{deluxetable}

\section{SEARCH AND IDENTIFICATION OF THE CANDIDATE COMPANIONS}\label{sec:discovery}

\subsection{Sample}
We searched for new binary systems comprising a low-mass component through common proper motion based on multi-epoch wide area near-infrared surveys. As a starting point for potential primary stars, we considered the NLTT \citep{salim_improved_2003} and LSPM \citep{lepine_nearby_2005} catalogs which include stars with proper motions larger than 0.18~mas~yr$^{-1}$ and 0.15~mas~yr$^{-1}$, respectively. We retained only the stars with a relative proper motion measurement error of less than 30\%. The large and precise proper motions of these stars are useful to limit the number of false positives, i.e., common proper motion between two sources resulting from chance or measurement errors.
To calculate the proper motion of the sources in the vicinity of these target stars, we compared the 2MASS PSC data \citep{cutri_2mass_2003} to data from the SDSS \citep{adelman-mccarthy_sixth_2008} and/or the SIMP \citep{artigau_simp:_2009} surveys. Of the roughly 60,000 stars included in the NLTT and LSPM catalogs, about 25,000 have accurate proper motion measurements and were observed by SIMP or SDSS; those constitute our initial sample. Target baseline's between the SDSS/SIMP and 2MASS observations span from 1 to 11 years, and the typical uncertainty of the calculated proper motions was 40~mas~yr$^{-1}$.

\subsection{Identifying Companions}
For each target star, we calculated the proper motion of all sources within a radius of 4\arcmin\ and selected any source with a proper motion within 40~mas~yr$^{-1}$ of the target star as a candidate companion. Visual inspection was performed for all candidates to reject false detections of artifacts that are identified within catalogs as point sources. The available photometry from the 2MASS PSC catalog was then used to estimate a spectral type for each candidate companion (and target star when unknown). We also estimated a photomeric distance and its uncertainty based on the $J$-band magnitude and the spectral type--magnitude relation of \cite{hawley_characterization_2002} when the information was not available in the literature. We required that the estimated distances of the target and candidate companion agree within the estimated uncertainties. Following this procedure, we identified 29 pairs of potential binary systems. We obtained imaging and spectroscopic follow-up observations of most of these candidates to improve their proper motion measurements and their spectral type determinations. We then applied a more detailed analysis, described below, to make a better assessment of their nature. Following this analysis, we found that 6 candidates systems are not comoving while 23 systems were found to be likely bound. These are the systems to be discussed in this paper (see Table~\ref{photo}).

Five of the 23 candidate pairs were found by other teams and reported as binary systems after we performed our initial search. The primary of these systems are NLTT~2274 (M4+M9.5), and NLTT~33793 (K4+M7.5), both found by \citet{faherty_brown_2010}, NLTT~20640 \citep[M4+L0]{zhang_discovery_2010}, NLTT~251 \citep[M7+M9+T5]{burgasser_discovery_2012}, and HD234344 \cite[K7,M7]{mason_2001_2001}.  In addition, shortly prior to the submission of this manuscript, a new study by \cite{deacon_wide_2014} reported the discovery of 4 other of our systems : NLTT 4558 (G5+L1+T3), NLTT56936 (K5+M8), NLTT26746 (M4+L4) and NLTT30510 (M2+M9.5).

\section{NEAR-INFRARED IMAGING}\label{sec:imaging}

\subsection{Observations}
We obtained $J$-band observations of 14 candidate companions between 2012 November and 2013 May using the Observatoire du Mont M\'egantic (OMM) wide field near-infrared camera, CPAPIR \citep{artigau_cpapir:_2004}, installed on the OMM 1.6~m telescope, and the OMM near-infrared spectro-imager SIMON \citep{albert_recherche_2006} installed on the CTIO 1.5~m telescope. The field of view (FOV) of CPAPIR at the OMM is 30\arcmin$\times$30\arcmin\ with a pixel scale of 0.89\arcsec~pixel$^{-1}$, while the FOV of SIMON at the Cerro Tololo Inter-American Observatory (CTIO) 1.5~m is 8.4\arcmin$\times$8.4\arcmin\ with a pixel scale of 0.49\arcsec~pixel$^{-1}$. Typically, the observations consisted of 15 individual exposures of 3 to 30~s, with a dither step of 2\arcmin\ between them. These observations provide a total time baseline of up to 15 years with respect to 2MASS for proper motion measurements.

\subsection{Data Reduction}
The reduction of the imaging data followed standard procedures. The images of SIMON and CPAPIR were sky subtracted using a median sky image constructed from the entire data set taken over the night in the $J$ filter. The images were flat-fielded using a flat image derived from \textit{on} and \textit{off} dome images taken at the beginning or the end of the nights. Individual images astrometry was performed by cross-correlating point sources in the field with the 2MASS catalog. All images were then median-combined into a single science frame that was used for astrometric measurements.

\subsection{Results}

We measured the positions of the sources in our CPAPIR and SIMON follow-up imaging by finding all of the stars in the image and then fitting a 2D Gaussian over each one.  These positions have been compared to those of all the objects in the 2MASS catalog that are up to 10\arcmin\ away from the target and the corresponding sources were associated. We determined the astrometric errors by computing the standard deviation of the difference between the positions in the CPAPIR/2MASS images and the corresponding positions in the 2MASS catalog propagated to the CPAPIR/SIMON epoch according to the proper motion listed in the NOMAD catalog \citep{Zacharias_NOMAD_2014}; this was done for reference stars of brightness comparable to our targets. We determined the source positions in the SIMP data, also obtained with CPAPIR, in the same manner; those measurements are also given in Table~\ref{radec} as they have not yet been published elsewhere.

\begin{deluxetable}{lrcrcclc}  
\tabletypesize{\tiny}
\tablewidth{0pt}
\tablecolumns{5}
\tablecaption{Astrometric measurements based on our observations\label{radec}}

\tablehead{   
  \colhead{Name} &
  \colhead{RA}  &
  \colhead{Uncertainty in RA}  &
  \colhead{DEC} &
   \colhead{Uncertainty in DEC} &
  \colhead{Epoch}&
  \colhead{Source} &
  \\
   &\colhead{(deg)}
   &\colhead{(mas)}
   &\colhead{(deg)}
 &  \colhead{(mas)}
 &  
  \colhead{(Reduced JD)}&
  &

}

\startdata
2M1043-1706A &160.84974     &     340      &          -17.10118     &     400  &       54168.75  &      SIMP &  \\
             &160.84956     &     560      &          -17.10128     &     670  &       56345.70  &     SIMON &  \\
             &160.84953   	&  	  130      &          -17.10125 	&     130  &       56259.97  &    CPAPIR &  \\\hline
2M1043-1706B &160.85452     &     340      &          -17.10234     &     400  &       54168.75  &      SIMP &  \\
            & 160.85435     &     220      &          -17.10245     &     110  &       56345.70  &     SIMON &  \\
             &160.85433 	&  	  250      &          -17.10240 	&     130  &       56259.97   &    CPAPIR &  \\\hline
             
2M1646+5019 &251.54744     &     190      &           50.33065     &     190 &       55232.88 &    SIMP &  \\
            &251.54735 	   &     400      &           50.33103     &     180 &       56330.86 &    CPAPIR &  \\\hline
            
LSPM1259+1001 & 	194.92540    &     860    &           10.02808     &     720   &       55232.91    &      SIMP &  \\
              &		194.92522	 &     680    &           10.02804 	   &     270   &       56330.82    &    CPAPIR &  \\\hline
2M1259+1001   &     194.92335    &     860    &           10.02741     &     720   &       55232.91    &    SIMP &  \\
              & 	194.92339    &     320    & 		  10.02728     & 	 220   &  	   53900.55    &    CPAPIR &  \\
              & 	194.92327    &     380    &           10.02737 	   &     270   &       56330.82    &    CPAPIR &  \\\hline  
              
LSPMJ1441+1856 &          220.34193     &     380    &           18.94546     &    390 &       55232.92  &      SIMP &  \\
               &		  220.34189 	&     160    &           18.94529 	  &    130 &       56330.89  &    CPAPIR &  \\\hline
2M1441+1856    &		  220.35673     &     380    &           18.94627     &    390 &       55232.92 &       SIMP &  \\
               & 		  220.37645 	&     260    &           19.09413     &    290 &       56330.89  &    CPAPIR &  \\\hline
            
NLTT182 	&  1.42300     &     150    &           6.44028      &     170 &       53980.73 & SIMP &  \\\hline
2M0005+0626 &  1.42296     &     150    &           6.44028      &     170 &       53980.73 & SIMP &  \\
            &  1.43611     &     360    &           6.51388      &     410  &       56619.50 &     SIMON &  \\\hline
            
NLTT20640   & 134.65331     &     310    &           27.18477    &     350  &       55144.94   &      SIMP &  \\
            & 134.65344		&     120    &           27.18451 	 &     170  &       56329.48   &    CPAPIR &  \\\hline
2M0858+2710 & 134.65424     &     310    &           27.18052    &     350  &       55144.94   &      SIMP &  \\
            & 134.67524 	&     300    &           27.31993 	 &     340  &       56329.48   &     CPAPIR &  \\\hline
            
NLTT33793   & 200.26834     &     190    &           10.02453   &     180  &       53838.67   &      SIMP &  \\
            & 200.26833 	&     360    &           10.02452 	&     280  &       56330.83  &    CPAPIR &  \\\hline
2M1320+0957 & 200.17290     &     190    &           9.96377    &     180  &       53838.67  &      SIMP &  \\
            & 200.17247 	&     110    &           9.96350 	&     90   &       56330.83 &    CPAPIR &  \\\hline
            
NLTT36369 	& 212.07912     &     360    &           37.14177     &     250 &       55232.91 &      SIMP &  \\
          	& 212.08075 	&     330    &           37.14089 	  &     870 &       56330.80  &    CPAPIR &  \\\hline
2M1408+3708 & 212.07911     &     360    &           37.14177     &     250 &       55232.91    &      SIMP &  \\
            & 212.08084 	&     380    &           37.14090 	  &     390 &       56330.80  &    CPAPIR &  \\\hline
            
NLTT41701 & 239.78411     &     470    &           37.59615     &    400 &       55232.89 &      SIMP &  \\
          & 239.78417 	  &     150    &           37.59600     &    160 &       56430.58 &    CPAPIR &  \\\hline
2M1559+3735 &239.78101    &    	470    &           37.59051     &    400 &       55232.89  &      SIMP &  \\
            &239.78105 	  &     120    &           37.59038     &    100  &      56430.58 &    CPAPIR &  \\\hline
            
2M0122+0331 & 20.57078     &   210     &           3.52294     &     150 &       54036.68 &      SIMP &  \\\hline

NLTT56936 	& 352.02213     &     130    &           4.97526     &     170 &       53986.70 &      SIMP &  \\
          	& 351.95329     &     290    &           4.85733     &     250 &       56572.54  &     SIMON &  \\\hline
2M2327+0450 & 351.95691     &     130    &           4.84984     &     170 &       53986.70 &      SIMP &  \\
            & 351.95778     &     70     &           4.85024     &     290 &       56572.54   &     SIMON &  \\\hline
           
2M0026+1704 &6.74936     &    220    &           17.07940     &     280 &       53994.73 &      SIMP &  \\\hline

BD-06813 	& 61.26145     &     470     &          -6.00778     &     470 &       56349.54  &     SIMON &  \\
         	& 61.26139	   &     140     &          -6.00779 	 &     190 &       56258.78 &    CPAPIR &  \\\hline
2M0405-0600 & 61.25892     &     180     &          -6.01190     &     180 &       56349.54 &     SIMON &  \\
            & 61.25890 	   &     110     &          -6.01187	 &     110 &       56258.78  &    CPAPIR &  \\\hline
            
NLTT2274 	& 10.48038     &     360    &           13.68728     &     220 &       56619.54   &     SIMON &  \\\hline
2M0041+1341 & 10.47664     &     140    &           13.69254     &     70  &       56619.54  &     SIMON &  \\\hline

2M2355+1754 & 358.89345     &    360    &           17.90071     &     140 &       56619.54 &     SIMON &  \\\hline

NLTT687 	& 3.50881     &     1080    &          -18.24962     &     2160 &       56547.70   &     SIMON &  \\\hline
2M0013-1816 & 3.49509     &     1080    &          -18.27954     &     2160 &       56547.70 &     SIMON &  \\\hline

2M2259+0806 & 344.81841   &     140     &           8.11571      &     140  &       56570.54  &     SIMON &  \\\hline

2M0006-0852 & 1.70457     &     250     &          -8.88072      &     220  &       56547.75 &     SIMON &  \\\hline

LSPMJ1021+3704 & 155.46781 &    310    &           37.07414 &     70  &       56329.67  &    CPAPIR &  \\\hline
2M1021+3704    & 155.47389 &    490    &           37.07081 &     190 &       56329.67      &    CPAPIR &  \\\hline

NLTT26746 	& 168.75463 &     130    &           16.11187 &     90  &       56329.69    &    CPAPIR &  \\\hline
2M1115+1607 & 168.75524 &     220    &           16.11687 &     150 &       56329.69   &    CPAPIR &  \\\hline

NLTT30510 	& 185.58421 &     180     &           36.73147 &     110  &       56329.73   &    CPAPIR &  \\\hline
2M1222+3643 & 185.57769 &     290     &           36.72998 &     300  &       56329.73 &    CPAPIR &  \\\hline

LSPMJ1236+3000 & 189.14888  &     120    &           30.00824 &   110 &       56329.71 &    CPAPIR &  \\\hline
2M1236+3000    & 189.15272  &     410    &           30.00832 &   210 &       56329.71    &    CPAPIR &  \\\hline

2M1202+4204 & 180.70753 &     580    &           42.08037 &    640 &       56330.73  &    CPAPIR &  \\\hline       

\enddata

\end{deluxetable}

\begin{figure}[ht]
\centering
\epsscale{1}
\plotone{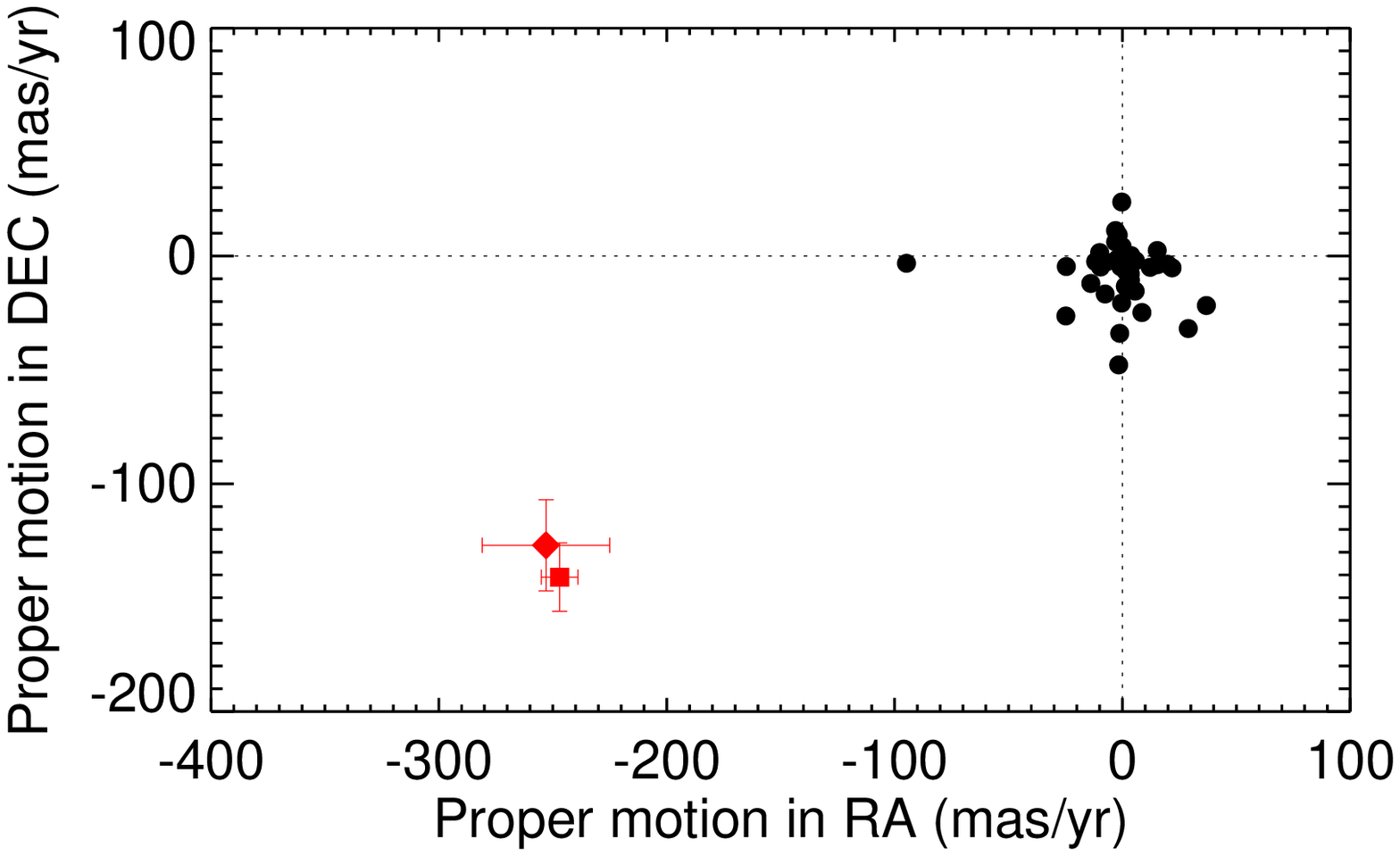}
\caption{\label{fig:pm} Proper motions for all objects within 10\arcmin\ of NLTT~26746 based on our follow-up imaging (2MASS, SDSS, CPAPIR, and WISE data). The proper motion of the candidate companion 2M1115+1607 is shown with a red diamond, while the proper motion of the primary star is shown with a red square. The two components of the candidate binary have proper motions consistent within the estimated uncertainties.}
\end{figure}

For each candidate companion, we combined the position measurements from our follow-up imaging, 2MASS, SDSS, SIMP, and WISE to compute their proper motion using a weighted linear regression between the positions of the object and the measurement epochs. We adopted the uncertainty of the slope parameter as our uncertainty on the proper motion. The proper motions for all of our candidates, corrected for the parallax estimated from their photometric distance (see below), are compiled in Table~\ref{spt}. As an example, our calculated proper motions for all stars within 10\arcmin\ of NLTT~26746 are plotted on Figure~\ref{fig:pm} ; the common proper motion of our candidate companion with the primary is clearly seen.

\section{SPECTROSCOPY}\label{sec:spectro}
\subsection{Optical spectroscopy}

\begin{deluxetable}{lccccccc}  
\tabletypesize{\tiny}
\tablewidth{0pt}
\tablecolumns{7}
\tablecaption{GMOS and GNIRS Spectroscopic Observations\label{obs_gmos}}
\tablehead{   
  \colhead{System} &
  \colhead{Component} &  
  \colhead{Obs. Date} &
  \colhead{Instrument} &
  \colhead{Number of exposures} &
  \colhead{Exp. time}&
  \colhead{Standard} &
  \\
     &
  &  
 \colhead{YYYY-MM-DD} &
  &
   &
  \colhead{(s)}

}
\startdata

NLTT251& Both & 2008-06-14          &          GMOS-S &    6 &      600. &         EG 274 \\\hline
NLTT 687 &  Primary &          2008-06-15 &          GMOS-S &    6 &      60.4 &         EG 274 \\
       & Companion &          2008-06-15 &          GMOS-S &    6 &      600. &         EG 274 \\\hline
NLTT 2274 & Primary &           2008-08-05 &          GMOS-S &    6 &      60.4 &       LTT 9239\\
          & Companion &           2008-08-05 &          GMOS-S &    3 &      900. &       LTT 9239 \\\hline
BD-06 813 & Primary &           2008-01-28 &          GMOS-S &    6 &      60.4 &         EG 274\\
         & Companion &          2008-01-28 &          GMOS-S &    6 &      600. &         EG 274\\\hline
NLTT 20640 & Both &          2008-02-11 &          GMOS-N &    6 &      599. &           HZ44\\\hline
LSPM J1021+3704 & Both &          2008-02-27 &          GMOS-N &    2 &      599. &           HZ44 \\\hline
2M1043-1706 &  Primary&        2008-01-28 &          GMOS-S &    6 &      600. &           HZ44 \\\hline
NLTT 26746 &  Primary &        2009-01-01 &          GMOS-S &    3 &      900. &       LTT 9239\\\hline
NLTT 29392 &  Both &        2008-02-11 &          GMOS-N &    6 &      599. &           HZ44 \\\hline
NLTT 36369 & Both &2008-04-14          &          GMOS-N &    9 &      599. &           HZ44 \\\hline
LSPM J1441+1856& Companion &          2007-08-06 &          GMOS-S &   10 &      600. &       LTT 9239 \\\hline
NLTT 41701 & Primary &         2008-03-15 &          GMOS-N &    6 &      9.99 &           HZ44 \\
           & Companion &         2008-03-15 &          GMOS-N &    9 &      599. &           HZ44\\\hline
HD 234344 & Primary &          2008-03-02 &          GMOS-N &    6 &      29.9 &           HZ44 \\
          & Companion &          2008-03-02 &          GMOS-N &    9 &      419. &           HZ44\\\hline
HD 217246 & Companion &          2007-06-24 &          GMOS-S &    6 &      600. &        LTT9239\\\hline
NLTT 56936 & Companion &          2008-12-27 &          GMOS-S &    3 &      900. &        LTT9239\\\hline
TYC 1725-344-1 & Companion &          2007-06-21 &          GMOS-S &    6 &      600. &        LTT9239\\\hline
NLTT 182 & Primary &   2008-06-30 &          GMOS-S &    6 &      600. &         EG 274\\\hline
2M1043-1706 & Companion  &        2007-04-01 &  GNIRS&  8 &      120. &HD~137873\\\hline
2M1115+1607 & Companion   &      2007-01-08 &  GNIRS& 11 &      60.0&HD~137873\\\hline
2M1259+1001 &  Companion   &     2007-01-08 &  GNIRS& 15 &      60.0&HD~137873 \\\hline

\enddata

\end{deluxetable}
We obtained optical spectroscopy follow-up observations of all candidates using the Gemini Multi-Object Spectrographs (GMOS) \citep{hook_gemini-north_2004} at the Gemini South and North telescopes during semesters 2008A and 2008B (programs GN-2008A-Q-80, GN-2008A-Q-20, GS-2008A-Q-12, GS-2008B-Q-34, and GN-2008B-Q-107). 
We used a 1\arcsec\ wide slit, the R400 grating with OG515 blocking filter, and 2--pixel binning in both the spatial and the spectral direction. For each target, we obtained exposures with three central wavelength settings, 790 nm, 800 nm and 810 nm, to cover the small gaps between the three GMOS detectors. The exposure times range from 10s to 600s (see Table~\ref{obs_gmos}). We did not move the target in the slit during the observations. The resulting spectra have a resolving power of $R$ $\sim$850 and cover the wavelength range from 650~nm to 1000~nm. The typical seeing during the observations was 0.75\arcsec--1.0\arcsec. Standard calibrations were obtained for each science data set.  We used the calibration data of the G subdwarf LTT9239, the DA white dwarf EG 274 and the B0 star HZ44 for telluric and instrumental transmission calibration.
Our main goal was to characterize the candidate companions, but given the short additional time needed, we also acquired a spectrum of the primary star during a visit to a given system. When the primary was not too bright to saturate the detector in an exposure of a few hundred seconds, both the primary and the companion were put in the slit simultaneously, otherwise a shorter observation of the primary was obtained after observation of the companion. Due to various issues during the execution of the observations, we were unable to procure data for all of the systems. In all, we obtained a spectrum for 14 candidate companions out for 23 and 10 candidate primaries out 23. A log of the observations is given in Table~\ref{obs_gmos}.

\begin{figure*}[p] 
\epsscale{2}
\plotone{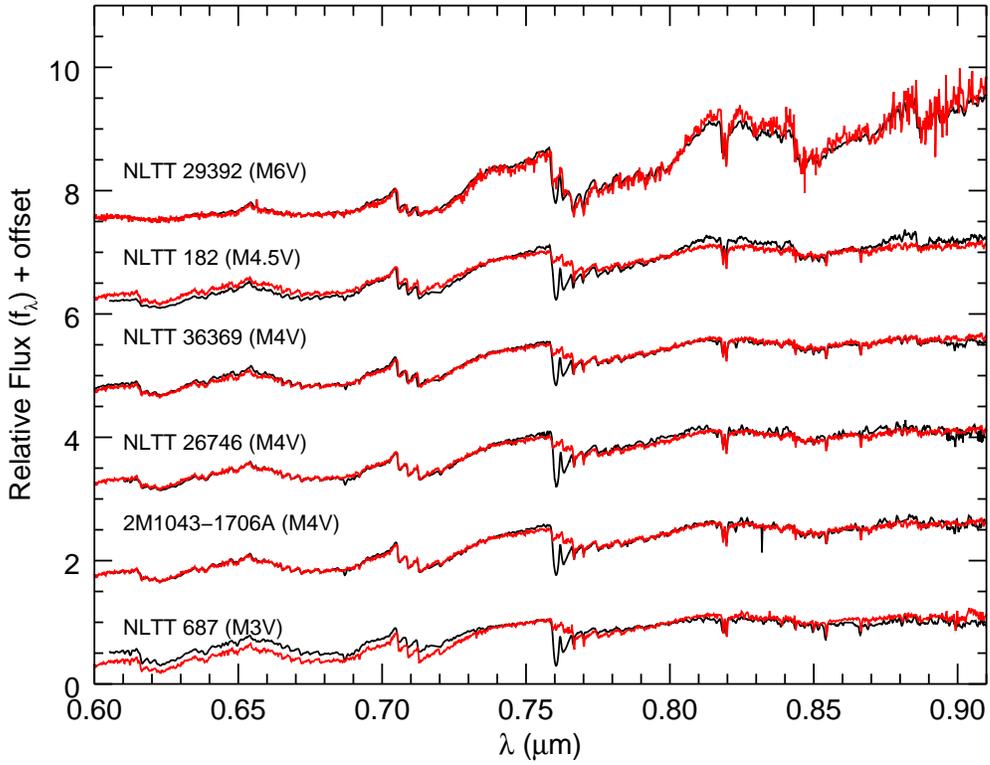}
\caption{GMOS spectra of our 6 candidate primaries for which a spectral type was not available in the literature, displayed with a resolution of R $\sim$ 850. The spectra are normalized at 0.75~$\mu$m. As a comparison, we display in red the SDSS template spectra from \cite{hawley_characterization_2002} for the appropriate spectral type convolved to the same resolution. The feature in our spectra at 0.76$\mu$m is telluric absorption that we did not correct. } 
\label{fig:gmos_prim}
\end{figure*}

\begin{figure*}[p] 
\epsscale{2}
\plotone{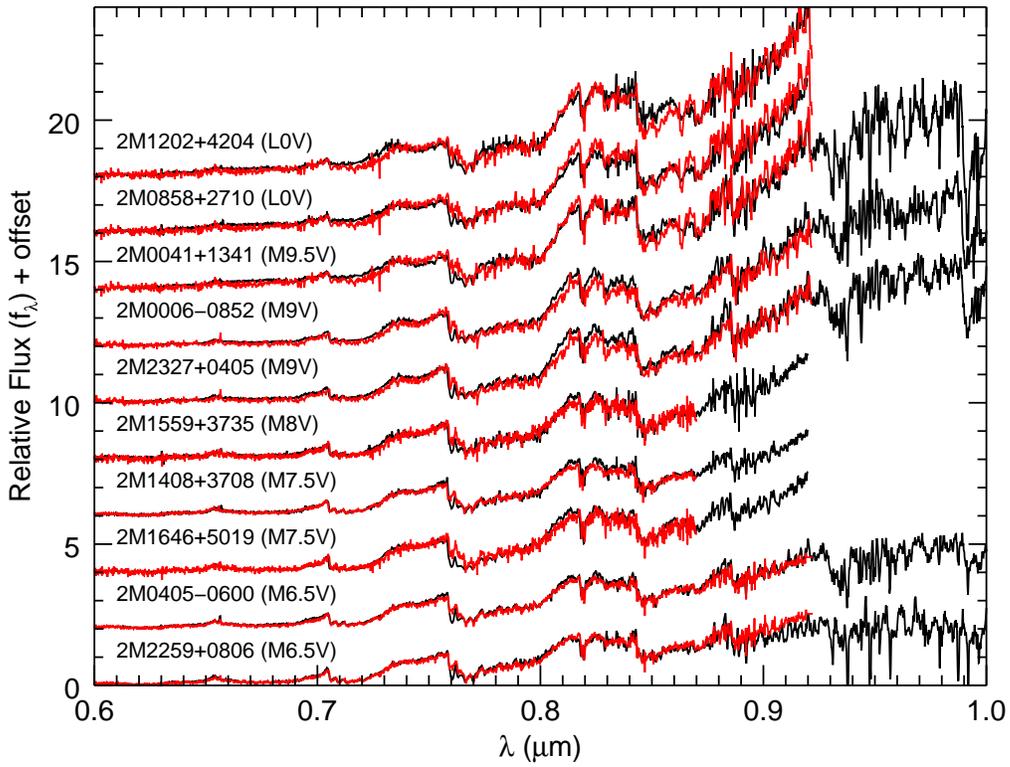}
\caption{GMOS spectra of 10 of our candidate companions, displayed with a resolution of R $\sim$ 850. The spectra are normalized at  0.75~$\mu$m. As a comparison, we display in red the SDSS template spectra from \cite{hawley_characterization_2002} for the appropriate spectral type convolved to the same resolution. The feature in our spectra at 0.76$\mu$m is telluric absorption that we did not correct. Different observation setups account for the different wavelength range.}
\label{fig:gmos_sec}
\end{figure*}

\begin{figure*}[p] 
\epsscale{2}
\plotone{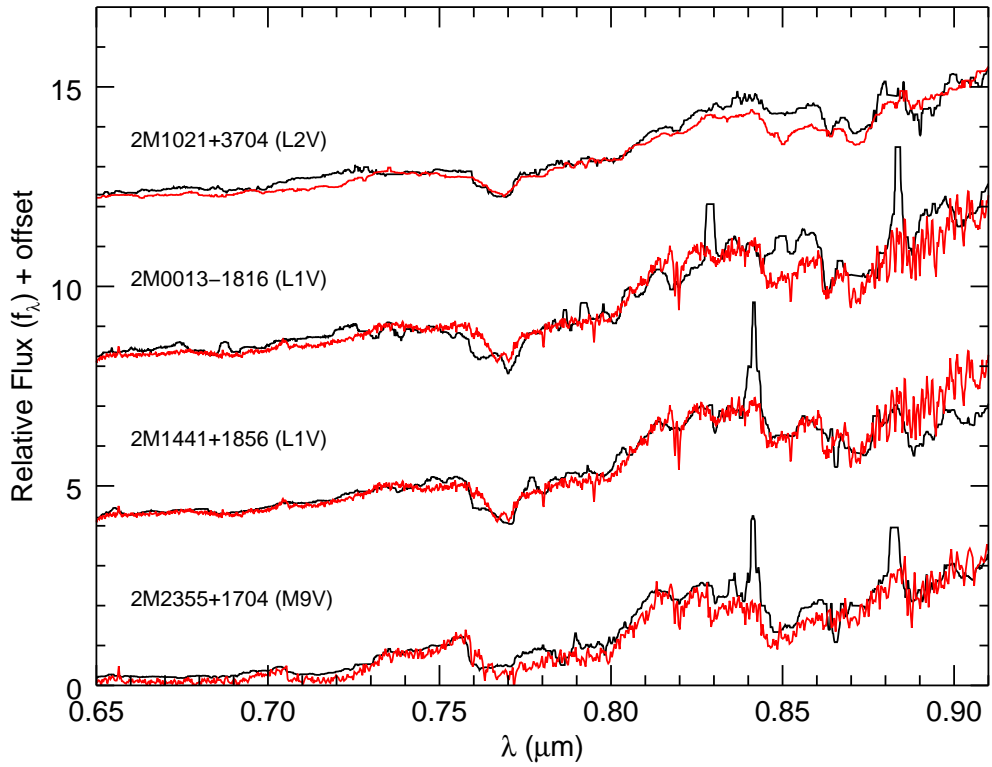}
\caption{GMOS spectra of 4 of our candidate companions, displayed with a degraded resolution of R $\sim$ 400 as these candidates were fainter and a lower signal-to-noise ratio was reached at full resolution. There is fringing in the data leading to a significant residual signal for these faint targets. The spectra are normalized at 0.75~$\mu$m. As a comparison, we display in red the SDSS template spectra from \cite{hawley_characterization_2002} for the appropriate spectral type. }
\label{fig:gmos_deg}
\end{figure*}

The GMOS spectroscopy data were reduced using a custom Interactive Data Language ({\em IDL}) routine. First, we corrected the individual exposures for the bias and flat field. We then extracted the spectral trace using a Moffat function extraction profile, allowing for a linear sky solution. 
We subsequently constructed a quadratic wavelength solution using calibration arc lamps, and corrected the spectra for wavelength-dependent instrumental transmission using a standard spectrum. The reference spectroscopic spectra were taken from \cite{hamuy_southern_1994} and \cite{massey_spectrophotometric_1988}.
In the last step, we combined the individual spectra extracted from each of the 3 detector section, and finally combined all exposures together for a given object. The measurement uncertainties were estimated by extracting an empty sky region near the spectral trace, using the same method. All these spectra are normalized at 0.75~$\mu$m. Figure~\ref{fig:gmos_prim} shows the GMOS spectra for 6 candidate primaries for which the spectral types were not published in the literature before this study. GMOS spectra at $R$ $\sim$ 850 of ten candidates are shown in Figure~\ref{fig:gmos_sec} and the remaining four are shown in Figure~\ref{fig:gmos_deg} with $R$ $\sim$ 400. The latter figure present the smoothed spectra our four faintest targets observed with GMOS. In all cases, a residual signal caused by an fringing effect from the detector is present. However, this effect affects the quality of our data significantly only for our faintest targets but even with this fringing, we are still able to identify the spectral type of the objects.

\begin{figure*}[p] 
\epsscale{2}
\plotone{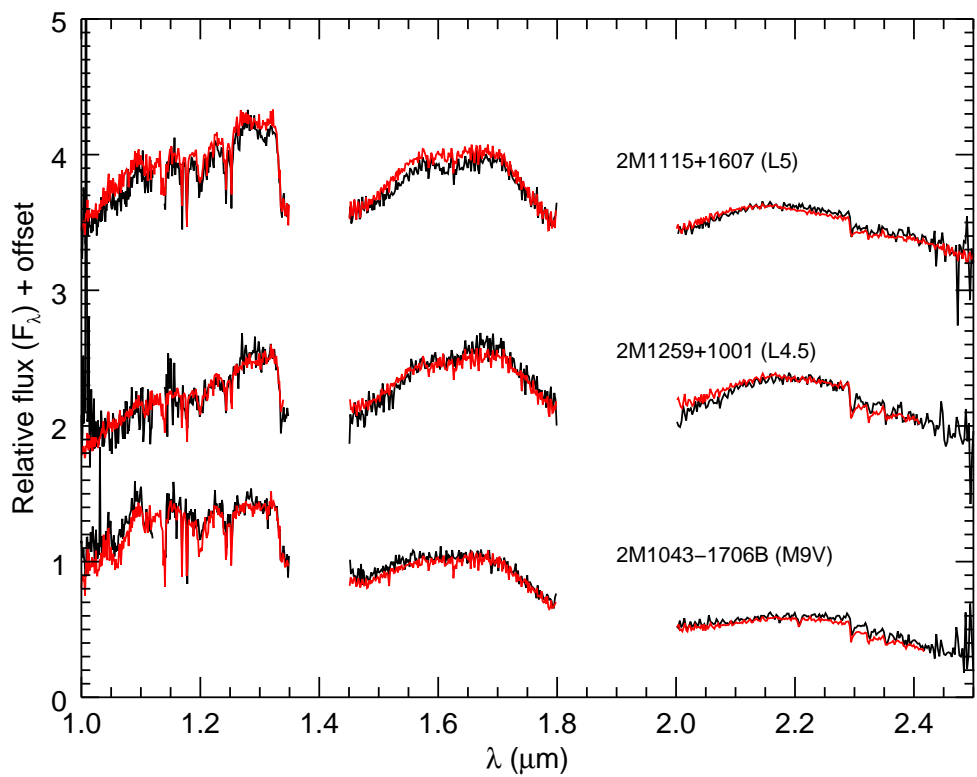}
\caption{GNIRS spectra of 3 of our candidate companions, displayed with a spectral resolution of R $\sim$ 1700. The spectra are normalized at 1.1~$\mu$m.As a comparison, we display in red the spectra from SPEX IRTF Spectral Library \citep{cushing_infrared_2005,rayner_infrared_2009} for the appropriate spectral type.}
\label{fig:gnirs}
\end{figure*}

\subsection{Near-Infrared Spectroscopy}

Three of our candidate companions, 2M1043-1706B, 2M1115+1607, and 2M1259+1001, were identified on their own as brown dwarf candidates in a separate search for high proper motion brown dwarfs in the SIMP data \citep[Robert et al., in preparation]{artigau_simp:_2009}, but have not yet been published. As other candidates identified in that search for brown dwarfs, they were followed-up using near-infrared spectroscopy with GNIRS at the Gemini South telescope in semester 2007A (GS-2007A-Q-28). We used a 0.3\arcsec\ wide slit, the 32~l~mm$^{-1}$ grating, the short camera, and the cross-dispersed mode for a resolving power of $R$ $\sim$ 1700 and a spectral coverage from 0.9~$\mu$m to 2.5~$\mu$m. Observatory standard calibrations were obtained with each observation. For telluric and instrumental transmission calibration, the A0 star HD~137873 was observed shortly before/after each observation. The observations were obtained with a typical ABBA dither pattern along the slit with individual exposures of 60 to 120s with 8 to 15 exposures. The details of the observations are shown in Table ~\ref{obs_gmos}. 

The reduction of the spectra was made using custom ({\em IDL}) routines as described in \cite{delorme_cfbds_2008}. First, successive image pairs were subtracted to remove the sky emission and divided by a median combined spectral flat. They were then corrected for both spectral and spatial distortions. To extract the spectra, a positive and a negative extraction box that matched the trace profile was used. An argon arc  lamp spectrum was taken at the end of each sequence, which was used as a first wavelength calibration. The wavelength scale was then more precisely adjusted to the atmospheric OH lines \citep{rousselot_night-sky_2000}. Finally, the spectra for each target extracted from image pairs were normalized and their median was taken to make the final spectra.  The GNIRS spectra of the three companions are shown on Figure~\ref{fig:gnirs} along with a spectra from the SPEX IRTF Spectral Library \citep{cushing_infrared_2005,rayner_infrared_2009}\footnote{\url{http://irtfweb.ifa.hawaii.edu/~spex/IRTF_Spectral_Library}} corresponding to the same spectral type as determined below. 

\subsection{Results}

We determined the spectral type of all objects observed with GMOS from spectral indices, and then confirmed them by visual comparison with templates. For each spectrum we computed the following spectral indices: PC3, PC4, and PC5 as defined by \cite{martin_spectroscopy_1996} and \cite{martin_spectroscopic_1999}, TiO5 and VO from \cite{cruz_meeting_2002}, CaH, Ti I, and Ca II from \cite{kirkpatrick_standard_1991}, and VO 2 and TiO7 defined by \cite{lepine_spectroscopy_2003}. Each of these indices constitutes a good spectral type indicator in some restricted spectral type interval, but altogether they cover the range from K5 to L6 dwarfs. The index values--spectral type relations and regimes of validity were taken from their respective papers.
The measured indices are compiled in Table~\ref{indices}, along with the corresponding spectral types. The latter are averages of the spectral types obtained from all the indices that are valid for a given type. The few blanks in the table are caused by the shorter spectral coverage of some spectroscopic observations. We visually compared our spectra to templates from the Pickles Atlas \citep{pickles_stellar_1998} or to SDSS spectra of M and L dwarfs from \citet{hawley_characterization_2002} to confirm spectral typing. The spectral types of all companions observed with GMOS range from M6.5 to L1.

\begin{deluxetable}{lccccccccccc}  
\tabletypesize{\tiny}

\tablecolumns{12}
\tablecaption{Spectral Indices\label{indices}}

\tablehead{   
  \colhead{Object} &
  \colhead{PC3$^{\textrm{a}}$} &
  \colhead{PC4$^{\textrm{a}}$} &
  \colhead{PC5$^{\textrm{a}}$}&
  \colhead{TiO5$^{\textrm{b}}$} &
  \colhead{VO$^{\textrm{b}}$}&
  \colhead{CaH$^{\textrm{c}}$} &
  \colhead{Ti I$^{\textrm{c}}$}&
  \colhead{Ca II$^{\textrm{c}}$} &
  \colhead{VO2$^{\textrm{d}}$}&
  \colhead{TiO7$^{\textrm{d}}$} &
  \colhead{Spectral Type $^{\textrm{e}}$}

}
\startdata
 NLTT251 &      1.3695 &      1.6816 &      1.7827 &     0.19484 &      2.1168 &      1.5603 &      1.0704 &      1.0764 &     0.56026 &     0.65387         &M6 $\pm$ 0.5  \\\hline
2M0006-0852 &      2.3435 &      3.8938 &      4.8902 &     0.34909 &      2.3607 &      1.5009 &      1.0326 &      1.0677 &     0.32328 &     0.46659         &M9 $\pm$ 0.5 \\\hline
NLTT687 &      1.0306 &      1.0248 &     0.93255 &     0.58126 &      2.0021 &      1.1914 &      1.0622 &      1.2380 &     0.89734 &     0.92856         &M3 $\pm$ 0.5  \\\hline
2M0013-1816 &      2.8484 &      6.2637 &      7.3279 &     0.96672 &      2.3180 &      1.4236 &      1.0617 &      1.1923 &     0.60007 &     0.95069         &L1 $\pm$ 0.5 \\\hline

NLTT2274 &      1.1037 &      1.1221 &      1.0918 &     0.40346 &      1.9970 &      1.2715 &      1.0803 &      1.1374 &     0.77698 &     0.85842         &M4 $\pm$ 0.5 \\\hline
2M0041+1341 &      2.7069 &      4.6043 &      5.7158 &     0.68651 &      2.5128 &      1.3602 &      1.0093 &      1.0645 &     0.36967 &     0.62286         &M9$\pm$ 0.5  \\\hline

2M0405-0600 &      1.6339 &      2.3019 &      2.4712 &     0.26199 &      2.2343 &      1.3667 &      1.1098 &      1.1253 &     0.42198 &     0.57595         &M6 $\pm$ 1  \\\hline

NLTT20640 &      1.1217 &      1.2361 &   \nodata       &     0.37900 &      2.0244 &      1.3055 &      1.0679 &      1.1134 &     0.72927 &     0.85573         &M4 $\pm$ 0.5 \\\hline
2M0858+2710 &      2.4959 &      4.7033 &   \nodata       &     0.64835 &      2.3871 &      1.2406 &     0.98143 &      1.0637 &     0.38629 &     0.65786         &L0 $\pm$ 1 \\\hline

LSPMJ1021+3704 &      1.0898 &      1.1628 &   \nodata       &     0.38845 &      1.9939 &      1.2819 &      1.0796 &      1.1583 &     0.75780 &     0.88121         &M4 $\pm$ 0.5 \\\hline
2M1021+3704 &      2.7920 &      5.3320 &  ... &      1.0516 &     0.46435 &     0.92451 &          ... &      1.0668 &     0.60661 &     0.66820         &L2  \\\hline

2M1043-1706A &      1.0665 &      1.1986 &      1.1118 &     0.41786 &      1.9828 &      1.2421 &     \nodata      &      1.1573 &     0.77439 &     0.87779         &M4 $\pm$ 0.5 \\\hline

NLTT26746 &      1.0896 &      1.0664 &      1.1495 &     0.44003 &      1.9733 &      1.2758 &      1.1094 &      1.1651 &     0.83645 &     0.85031         &M4 $\pm$ 0.5 \\\hline

NLTT29392 &      1.3948 &      2.0186 &   \nodata       &     0.21082 &      2.1644 &      1.4311 &      1.0827 &      1.0974 &     0.49018 &     0.65474         &M6 $\pm$ 0.5 \\\hline
2M1202+4204 &      2.7094 &      5.3242 &   \nodata       &     0.66899 &      2.4729 &      1.3948 &     0.96756 &      1.0718 &     0.39638 &     0.70797         &L0 $\pm$ 0.5  \\\hline

NLTT36369 &      1.0359 &      1.0841 &    \nodata      &     0.42531 &      2.0007 &      1.2596 &      1.0741 &      1.1489 &     0.80273 &     0.89932         &M4 $\pm$ 0.5 \\\hline
2M1408+3708 &      1.6805 &      2.5229 &    \nodata      &     0.22558 &      2.3573 &      1.4836 &      1.0556 &      1.1543 &     0.37181 &     0.57255         &M7 $\pm$ 0.5  \\\hline

2M1441+1856 &      2.7019 &      4.2794 &      2.7264 &     0.88903 &      2.2368 &      1.5198 &     0.75043 &      1.2276 &     0.50723 &     0.76976         &L1 $\pm$ 1 \\\hline

2M1559+3735 &      1.8393 &      2.9506 &    \nodata      &     0.27357 &      2.4630 &      1.3777 &      1.0318 &      1.1406 &     0.33012 &     0.55086         &M8 $\pm$ 0.5 \\\hline

2M1646+5019 &      2.0450 &      3.0980 &    \nodata      &     0.17044 &      2.1751 &      1.8641 &      1.0559 &      1.1482 &     0.39253 &     0.54864         &M7 $\pm$ 0.5 \\\hline

2M2259+0806 &      1.5610 &      1.8795 &      1.6930 &     0.14590 &      2.1279 &      1.5757 &      1.0905 &      1.2289 &     0.50888 &     0.61669         &M6 $\pm$ 0.5 \\\hline

2M2327+0405 &      2.1572 &      3.2263 &      3.8603 &     0.27870 &      2.4432 &      1.4433 &      1.0444 &      1.0839 &     0.33743 &     0.49614         &M9 $\pm$ 0.5 \\\hline

2M2355+1754 &      2.2927 &      3.7816 &      2.9846 &     0.42313 &      2.4953 &      1.3452 &      1.0084 &      1.1327 &     0.40961 &     0.76501         &M9 $\pm$ 0.5 \\\hline
 NLTT182 &      1.1426 &      1.2647 &      1.1793 &     0.34680 &      2.0167 &      1.3121 &      1.0885 &      1.1310 &     0.69285 &     0.79268         &M4.5 $\pm$ .5  \\\hline
\enddata
\tablenotetext{a}{ The spectral indice is from \cite{martin_spectroscopy_1996}}
\tablenotetext{b}{The spectral indice is from \cite{cruz_meeting_2002}}
\tablenotetext{c}{The spectral indice is from \cite{kirkpatrick_standard_1991}}
\tablenotetext{d}{The spectral indice is from \cite{lepine_spectroscopy_2003}}
\tablenotetext{e}{The spectral types shown in the last column are the median of the spectral types computed using the spectral indices presented in the table. }

\end{deluxetable}

We determined the spectral type of the three companions observed with GNIRS using the spectral indices defined by \cite{allers_near-infrared_2013}, which are based on the H$_2$O, H$_2$O-1, H$_2$O-2, and H$_2$OD indices and which are valid for M5V to L7V dwarfs. Then, we visually compared our spectra to the templates of M and L dwarfs from the SPEX IRTF Spectral Library. We determined a spectral type of M9.0$\pm$0.5 for 2M1043-1706B, L5 $\pm$ 1 for 2M1115+1607, and L4.5 $\pm$ 0.5 for 2M1259+1001; these are indicated in Table~\ref{spt}.

There are 19 objects for which we do not have GMOS nor GNIRS observations. Among these are 10 primaries and 2 companions for which a spectral type was known from the literature and we adopted those values. For the remaining seven, we estimated a spectral type from their 2MASS and SDSS colors based on the color--spectral type relations of \cite{sheppard_infrared_2009}. The spectral types adopted for all objects are summarized in Table~\ref{spt}. This table is divided in the two parts: the top half is for systems for which we have a spectrum for the companions while the lower half is for systems for which we did not have a spectrum for the companions.

For each object we estimated a photometric distance using their 2MASS $J$--band magnitude and the $M_J$--spectral type relations from \cite{hawley_characterization_2002} for late-K to M5 dwarfs, and from \citet{dupuy_hawaii_2012} for later type objects, M6 to L5. These distance estimates are given in Table~\ref{spt}.

\section{PROBABILITY OF RANDOM ALIGNMENT}\label{sec:prob}
After determining the proper motion, the spectral types and the distance for all components, we need to confirm that they form a bound pair. The probability of random alignment is the probability that from chance, in a search like ours, we would find a physically unrelated companion star having the same proper motion and photometric distance as the primary, within our uncertainties, and that is separated by less than the observed separation of our candidate. 
This is given by the probability of finding two stars of the relevant spectral types within a given separation of each other and at the same distance, times the probability of finding two stars of the relevant spectral types with the same proper motion, given that they are at a similar location on the sky. 

For the former probability, as our search targeted a fixed set of potential primary stars, we need only calculate the probability of finding a companion star close to those primaries. All of our candidate companions have spectral types in the M6--L5 range, thus this is the relevant range for statistical calculations. From the spatial density of M6-M8 dwarfs of 2.2 $\times$ 10$^{-3}$ pc$^{-3}$ per $I$-band magnitude interval determined by \citet{phan-bao_new_2003} and the spatial density of M8-L3.5 of $1.64\times10^{-3}$~pc$^{-3}$ per $J$-band magnitude interval determined by \citet{phan-bao_discovery_2008}, we calculated the overall spatial density of late-M dwarfs to early-L dwarfs (M6 to L3.5). We used this density to calculate the number of such stars in a spherical shell of radius equal to the distance from the Sun to a candidate and thickness given by our distance estimate uncertainty. Then, from this number we calculated, the average number of stars in a sky projected disk of radius equal to the separation of our candidate binary, assuming an isotropic distribution of stars. We multiplied this number by 25 000, the number of primary stars targeted by our search, and finally calculated the corresponding probability of detecting at least one object within this total search area.

For the latter probability, we used Monte Carlo calculations to determine the proper motion distribution of stars of a given spectral type and at the sky position of the candidate system based on the observed Galactic space velocity ($UVW$) distributions of stars in the solar neighborhood. The $UVW$ distributions were derived from the velocity dispersions given by \cite{mikami_absolute_1982} for F, G and K stars, by \cite{bochanski_sloan_2011} for M dwarfs, and by \cite{schmidt_colors_2010} for L dwarfs. We then determined the probability that a star of the spectral class of the primary and a star of the spectral class of the secondary, at the sky position of the system, would have proper motions larger than 0.1~mas~yr$^{-1}$ and consistent with each other within 2$\sigma$. 

We finally multiplied the above two probabilities to get the probability of random alignment; the results are given in Table~\ref{spt}. For all of our candidate systems, this probability is less than $1.3\times10^{-3}$, indicating that they most likely form a physical pairs.

\begin{deluxetable}{lcccccccccc}  

\tabletypesize{\tiny}
\tablewidth{0pt}
\tablecolumns{9}
\tablecaption{Derived parameters for the binary systems\label{spt}}

\tablehead{   
  \colhead{Name} &
  \colhead{Spectral type$^{\textrm{a}}$} &
  \colhead{$\mu_\alpha \cos \delta$ } &  
  \colhead{$\mu_\delta$ } &
  \colhead{Separation }&
  \colhead{Distance$^{\textrm{b}}$ } &
  \colhead{Prob $^{\textrm{c}}$} &  
  \colhead{P. sep.} &
    \colhead{Pos. angle} &
  \colhead{Mass $^{\textrm{d}}$} 

\\
   &
  &
   
  \colhead{(mas/yr)} &  
  \colhead{(mas/yr)} &
  \colhead{($\prime\prime$)}&
  \colhead{(pc)} &
  &
  \colhead{(AU)} &
    \colhead{(deg)} &
  \colhead{($M_{\odot}$)}  
  
}

\startdata
NLTT 251$^{\textrm{p}}$&M$6.0\pm0.5$&$-70\pm9$&$-311\pm6$&$27.4\pm0.6$&$25^{+7}_{-11}$&2.8 $\times 10^{-5}$&$850^{+84}_{-152}$&$113\pm1$& 0.102--0.133   \\
2M0006-0852&M$9.0\pm0.5$&$-59\pm18$&$-318\pm9$&&$37^{+5}_{-11}$&&&&0.079--0.085  \\\hline

NLTT 687&M$3.0\pm0.5$&  $-44\pm12$&$-173\pm5$&$118.1\pm0.6$&$42^{+26}_{-16}$&3.9 $\times 10^{-4}$ &$7400^{+1160}_{-1250}$&$203.1\pm0.3$&0.389--0.412     \\
2M0013-1816&L$1.0\pm0.5$& $-20\pm41$&$-212\pm37$&&$83^{+7}_{-27}$&&&&   0.072--0.078 \\\hline

NLTT 2274$^{\textrm{f}}$&M$4.0\pm0.5$&$-165\pm56$&$-155\pm2$&$23.2\pm0.8$&$23^{+14}_{-9}$&1.6 $\times 10^{-4}$&$725^{+230}_{-253}$&$325\pm2$& 0.207--0.272 \\\
2M0041+1341&M$9.5\pm0.5$&$-162\pm58$&$-153\pm13$&&$40^{+6}_{-12}$&&&&0.076--0.083\\\hline

BD-06 813  &K0$^{\textrm{o}}$&$31\pm8$&$-139\pm5$&$17.4\pm0.2$&$68\pm10^{\textrm{h}}$&7.6 $\times 10^{-4}$&$1340^{+102}_{-340}$&$211.6\pm0.5$&0.925--1.014\\
2M0405-0600&M$6.5\pm1$&$42\pm6$&$-117\pm10$&&$86^{+2}_{-37}$&&&&0.096--0.114\\\hline

NLTT 20640$^{\textrm{q}}$&M$4.0\pm0.5$&$103\pm5$&$-179\pm5$&$15.6\pm0.2$&$54^{+32}_{-20}$&2.2 $\times 10^{-5}$&$780^{+263}_{-270}$&$168.9\pm0.6$&0.207--0.272\\
2M0858+2710&L$0\pm1$&$104\pm7$&$-182\pm6$&&$48^{+3}_{-14}$&&&&  0.074--0.081   \\\hline

LSPM J1021+3704&M$4.0\pm0.5$ $^{\textrm{h}}$&$-132\pm9$&$-141\pm4$&$22.2\pm0.4$&$88^{+39}_{-25}$&1.3 $\times 10^{-3}$&$3000^{+650}_{-675}$&$125\pm1$&0.207--0.272\\
2M1021+3704&L$0\pm1$&$-131\pm8$&$-124\pm12$&&$93^{+13}_{-36}$&&&& 0.071--0.076 \\\hline

2M1043-1706A&M$4.0\pm0.5$&$-97\pm16$&$-140\pm22$&$17.1\pm0.1$&$42^{+26}_{-16}$&7.1 $\times 10^{-4}$&$1020^{+288}_{-306}$&$84\pm0.5$&0.207--0.272 \\
2M1043-1706B&M$9.0\pm0.5$&$-90\pm14$&$-141\pm22$&&$75^{+6}_{-23}$&&&& 0.079--0.085\\\hline

NLTT 26746$^{\textrm{u}}$ &M$4.0\pm0.5$&$-247\pm8$&$-141\pm2$&$18.0\pm0.2$&$36^{+22}_{-14}$&1.5 $\times 10^{-4}$&$660^{+216}_{-441}$& $6.5\pm0.7$&0.207--0.272   \\
2M1115+1607&L$5\pm1$ $^{\textrm{t}}$&$-253\pm28$&$-127\pm20$&&$37^{+3}_{-35}$&&&&0.056--0.073\\\hline

NLTT 29392 &M$6.0\pm0.5$&$-306\pm11$&$-300\pm17$&$7.3\pm0.1$&$33^{+2}_{-16}$&1.1 $\times 10^{-5}$&$310^{+16}_{-135}$&$42.0\pm0.8$&0.102--0.133\\
2M1202+4204&L$0.0\pm0.5$&$-292\pm6$&$-276\pm11$&&$38^{+2}_{-11}$&&&&0.074--0.081 \\\hline

LSPM J1259+1001&M$5\pm1^{\textrm{g}}$&$-142\pm15$&$22\pm9$&$7.65\pm0.08$&$42^{+44}_{-21}$&2.9 $\times 10^{-3}$&$345^{+196}_{-156}$&$250.9\pm0.6$&0.121--0.167 \\
2M1259+1001&L$4.5\pm0.5$ &$-146\pm5$&$29\pm7$&&$47^{+5}_{-18}$&&&&0.057-- 0.074\\\hline

NLTT 36369 &M$4.0\pm0.5$&$-247\pm8$&$80\pm1$&$7.9\pm0.1$&$55^{+33}_{-21}$&9.3 $\times 10^{-5}$&$590^{+162}_{-180}$&$122\pm1$&0.207--0.272 \\
2M1408+3708&M$7.5\pm0.5$&$-249\pm1$&$80\pm10$&&$84^{+4}_{-29}$&&&& 0.087--0.096 \\\hline

LSPM J1441+1856 &M$6^{\textrm{h}}$ &$-56\pm9$&$-154\pm9$&$51.1\pm0.2$&$56^{+37}_{-10}$& 8.7 $\times 10^{-4}$&$4110^{+1219}_{-1166}$&$86.4\pm0.2$&0.102--0.133\\
2M1441+1856&L$1\pm1$
&$-59\pm9$&$-185\pm12$&&$99^{+10}_{-34}$&&&&0.072--0.079\\\hline

NLTT 41701 &K$2^{\textrm{h}}$ &$42\pm12$&$-137\pm9$&$22.4\pm0.5$&$(61^{+61}_{-27})^{\rm{i}}$&8.1 $\times 10^{-4}$&$1735^{+690}_{-644}$&$204\pm1$&0.872--0.941  \\
2M1559+3735&M$8.0\pm0.5$ &$46\pm9$&$-152\pm14$&&$89^{+6}_{-29}$&&&& 0.084--0.090   \\\hline

HD 234344$^{\textrm{s}}$&K7$^{\textrm{h}}$ &$-129\pm29$&$394\pm11$&$69.3\pm0.3$&$31.6\pm0.9^{\rm{m}}$&2.4 $\times 10^{-4}$&$2700^{+158}_{-593}$&$146.2\pm0.3$&0.705--0.751 \\
2M1646+5019&M$7.5\pm0.5$&$-118 \pm4$&$400\pm7$&&$36^{+3}_{-14}$&&&&0.087--0.095 \\\hline

HD217246&G5$^{\textrm{h}}$  &$137\pm34$&$59\pm19$&$24.7\pm0.2$&$(68^{+70}_{-32})^{\rm{i}}$&3.6 $\times 10^{-4}$& $2340^{+1125}_{-1000}$ & $286.0\pm0.5$&0.821--0.930  \\
2M2259+0806&M$6.5\pm0.5$ &$122\pm6$&$69\pm5$&&$121^{+20}_{-51}$&&&& 0.096--0.113        \\\hline

NLTT 56936$^{\textrm{u}}$ &K2+K5$^{\textrm{n}}$ &$443\pm1$&$171\pm8$&$30.3\pm0.1$&$64\pm9^{\rm{k}}$&4.7 $\times 10^{-5}$&$1825^{+180}_{-390}$&$147.8\pm0.2$&1.46$\pm0.09$ $^{\textrm{n}}$\\
2M2327+04505&M$9\pm0.5$&$443\pm6$&$191\pm12$&&$57^{+3}_{-17}$&&&& 0.079--0.085\\\hline

TYC 1725-344-1&G5III$^{\textrm{h}}$  &$-65\pm21$&$-78\pm14$&$93.8\pm0.4$&$(40^{+28}_{-15})^{\rm{i}}$&2.8 $\times 10^{-4}$&$6700^{+1805}_{-2470}$&$149.0\pm0.23$&1.91$\pm0.11$ $^{\textrm{h}}$\\
2M2355+1754&M$9.0\pm0.5$ &$-69\pm16$&$-86\pm3$&&$103^{+10}_{-37}$&&&&0.079--0.085\\\hline\hline

NLTT 182 &M$4.5\pm0.5$&$213\pm6$&$-113\pm7$&$6.1\pm0.1$&$62^{+38}_{-23}$&1.4 $\times 10^{-4}$&$400^{+126}_{-135}$&$43\pm4$& 0.157--0.181    \\
2M0005+0626&L$0\pm1^{\textrm{g}}$&$205\pm7$&$-126\pm8$&&$69^{+5}_{-22}$&&&&0.079--0.085 \\\hline

LSPM J1236+3000 &M$6\pm1^{\textrm{g}}$&$140\pm5$&$-124\pm4$&$12.2\pm0.8$&$112^{+6}_{-52}$& 1.9 $\times 10^{-3}$&$1580^{+156}_{-624}$&$89\pm4$&0.102--0.133\\
2M1236+3000&M$9.0\pm1^{\textrm{g}}$&$162\pm27$&$-102\pm15$&&$126^{+18}_{-44}$&&&& 0.079--0.085\\\hline

HD 2292 &G5$^{\textrm{r}}$&$-143\pm15$&$-113\pm19$&$30.5\pm0.5$&$(42^{+28}_{-14})^{\rm{i}}$&2.6 $\times 10^{-4}$&$1200^{+180}_{-510}$&$138\pm1$&1.05-1.12   \\
2M0026+1704&M$9\pm1^{\textrm{g}}$&$-149\pm2$&$-128\pm41$&&$65^{+4}_{-20}$&&&&0.108--0.118\\\hline

NLTT 30510$^{\textrm{u}} $&M$3\pm1^{\textrm{g}}$&$269\pm12$&$-39\pm16$&$20.7\pm0.5$&$45^{+47}_{-23}$&1.9 $\times 10^{-4}$&$1635^{+648}_{-384}$& $254.2\pm0.9$&0.377--0.431\\
2M1222+3643& L0$^{\textrm{e}}$ & $272\pm18$&$-36\pm8$&&$70\pm10^{\rm{l}}$&&&&0.074--0.081\\\hline

NLTT 33793$^{\textrm{j}}$ &K$5^{\textrm{j}}$&$-210\pm64$&$-131\pm12$&$168.5\pm0.3$&$(38.1^{+2.6}_{-2.3})^{\rm{j}}$&5.4 $\times 10^{-5}$&$5780^{+2040}_{-1190}$&$311.7\pm0.1$& 0.724--0.782  \\
2M1320+0957&M8$\rm^{\textrm{j}}$ &$-222\pm38$&$-141\pm10$&&$36\pm3^{\textrm{j}}$&&&&0.087--0.096 \\\hline

NLTT 4558$^{\textrm{u}}$&G5$^{\textrm{m}}$&$22\pm13$&$-142\pm15$&$44.8\pm0.8$&$(58\pm3^{\rm{k}}$&3.5 $\times 10^{-5}$&$2222^{+308}_{-960}$& $324.2\pm1$&1.05-1.12 \\
2M0122+0331&L$2\pm1^{\textrm{g}}$&$45\pm10$&$-162\pm5$&&$43^{+13}_{-40}$&&&&0.071--0.076\\\hline

\enddata

\tablenotetext{a}{Spectral type are extracted from our GMOS or GNIRS spectra unless noted otherwise,$^{b}$ Distance computed using the spectral types and the relations from \cite{dupuy_hawaii_2012} and \cite{hawley_characterization_2002},  $^{c}$ Probability of random alignement,$^{d}$ Masses are evaluated using the mass--M$_J$ relation from the BT-Settl model of \cite{allard_bt-settl_2014}, $^{e}$\cite{zhang_ultra-cool_2009},$^{f}$ This system has been identified as a M4+L0 binary by \citet{faherty_brown_2010}. Their spectral types, proper motions, distance and sepration are consistent with the ones we found,$^{g}$ Spectral type calculated from i'-J \citep{sheppard_infrared_2009},$^{h}$ \cite{pickles_all-sky_2010},$^{i}$ \cite{ammons_n2k_2006}  has estimated a distance that matches to one obtained from the spectral type extracted from the optical GMOS spectrum, $^{j}$ This system has been identified as a M4+L0 binary by \citet{faherty_brown_2010},$^{k}$ Heliocentric distance from \cite{anderson_extended_2013},$^{l}$ Photometric distance from \cite{zhang_ultra-cool_2009},$^{m}$ Heliocentric distance from \cite{anderson_extended_2013},$^{n}$\cite{hrivnak_study_1995}, 
$^{o}$ The spectral type is from \cite{barney_supplementary_1951},$^{p}$ This system has been identified to be a hierarchical triple system with an M7.0$\pm0.5$ primary and an M8.5$\pm0.5$+T5$\pm1$ secondary by \cite{burgasser_discovery_2012}.  Their proper motions, distance and sepration are consistent with the ones we found,$^{q}$ This system has been identified as a M4+L0 binary by \cite{zhang_discovery_2010}. Their spectral types, proper motions, distance and sepration are consistent with the ones we found,$^{r}$ \cite{kharchenko_all-sky_2001},$^{s}$ This system has been identified as a K7+M7 binary by \cite{mason_2001_2001}. Their spectral types, proper motions, distance and sepration are consistent with the ones we found,$^{t}$\cite{zhang_ultra-cool_2009-1} found a spectral type of L1 by colors and has identified it as a brown dwarf candidate.,$^{u}$ This system has been identified by \cite{deacon_wide_2014},$^{v}$ The mass of NLTT 4558 has been calculated by \cite{porto_de_mello_photometric_2014}}

\end{deluxetable}

\section{DISCUSSION}\label{sec:discussion}

\afterpage{
\begin{figure}[t]
\centering
\epsscale{1}
\plotone{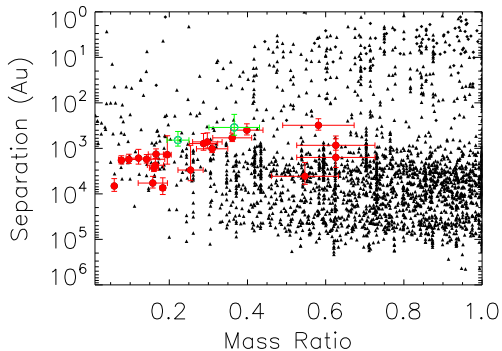}
\caption{\label{fig:sep_q}  Separation (in AU) versus mass ratio for our systems and for known binaries. The red circles represent our sample. The LSPMJ1259+1001 system and the NLTT26746 system, for which the companion is either a L4.5 or a L5 dwarf, are shown with the open green circle. Very low mass binaries from the Very Low Mass Binaries Archive\protect\footnotemark are shown as a diamond while the triangles represent binary systems from  \cite{gomes_two_2013} and references therein, \cite{janson_astralux_2012}, \cite{dhital_sloan_2010}, \cite{faherty_brown_2010} and references therein, \cite{raghavan_survey_2010}, \cite{metchev_search_2004},  \cite{reid_search_2001}, \cite{wilson_three_2001},\cite{mason_2001_2001-1}, \cite{reid_low-mass_1997}, \cite{fischer_multiplicity_1992}, and \cite{close_complete_1990}.}
\end{figure}

\footnotetext{See \url{http://www.vlmbinaries.org/} for an up-to-date census of the known very low mass binaries, maintained by N. Siegler, C. Gelino and A. Burgasser.}
}

We estimated the mass of the companions in our systems based on evolution models; this requires first estimating an age for the systems. The tangential velocity of our systems, which is an indicator of the population to which an object belongs, has been compared to the Besan\c{c}on Galactic model \citep{robin_synthetic_2003}.  All of our systems have tangential velocities consistent with being part of the thin disk. According to the Galactic model simulations, most of our targets have a $>$ 80\% probability to be younger than 3 to 5 Gyr. We also looked for sign of youth in our GNIRS spectra. We used the spectral indices defined by \citet{allers_near-infrared_2013}, already used to compute the spectral types, which are also sensitive to surface gravity. According to these indices, none of the three companions have low surface gravity. We then looked for H$\alpha$ emission, at 656.3~nm, which is indicative of the stellar activity and moderately correlated with age \citep{west_constraining_2008}. Visual inspection of the spectra and calculations of the H$\alpha$ equivalent widths indicate that 2M0405-0600 (M6.5), 2M1202+4204 (L0) and 2M1043B (M9) have equivalent widths higher than 1 $\textup{\AA}$, and are thus active. \citet{schmidt_colors_2010} had already identified that NLTT~29392B was an active L0 dwarf. However, this does not mean that they are particularly young as M dwarfs can be active during a few Gyr (see discussion in \cite{west_constraining_2008}) on the relatively weak constraint that H$\alpha$ provides for an individual object).  

In addition, we used the BANYAN II code \citep{gagne_banyan_2014} to test if any of our targets belong to a nearby young moving group. We found out that while most of our target have a higher probability of belonging to the field, one of them (NLTT182) has a 98.8\% probability of being a Beta Pictoris member, with a corresponding false positive rate of 4.5\%. This is quite interesting but by no means a proof of youth or true membership; the complete 3D kinematics would have to be confirmed, as well as youth from various indicators. We only have an optical spectrum of the primary star of this system, and we looked for gravity-sensitive spectroscopic indices in it \citep{kirkpatrick_infared_2000,cruz_meeting_2002,cruz_meeting_2007,reid_meeting_2008}, such as the Na-a index, the Na-b index, the K-a index, and the CrH-a index, to assess whether this object is young or not. However, these four indices fall close to the field scatter and the result of our analysis is mainly inconclusive. More observations will be necessary to firmly establish whether or not this system is young. For the purpose of this paper, due to the lack of solid evidence, we simply assume this system to be of the age of the field, as we do for all of our other systems. 

Based on the above analyses, we thus estimated that the ages of all of the systems are roughly 1--7~Gyr. We then evaluated the masses using the mass--M$_J$ relation from the BT-Settl model of \cite{allard_bt-settl_2014}, where the M$_J$ are obtained from our inferred spectral types using the relation of \cite{hawley_characterization_2002} and \cite{dupuy_hawaii_2012}. In the cases 
where the distance were not well defined, we decided to use the effective temperature
instead of the $J$ magnitude to find the masses.  The effective temperature for 
each target has been extracted from its spectral type according to the relations in  \cite{pecaut_intrinsic_2013}.  We note that masses infered in this manner are overestimated by about 10\% for objects with a G or early--K spectral types as compared to \cite{henry_mass_1993}.  The estimated masses for ages between 1 to 7 Gyr are given in Table~\ref{spt}. The uncertainties come from the propagation of the error of spectral types on the inferred magnitudes (or effective temperatures).

\begin{sloppypar}
Figure~\ref{fig:sep_q} shows the separation in AU of our systems as a function of their mass ratio, as compared with other known binaries. Some of our new systems, having low mass ratios, reach a sparsely populated regions of the diagram. In particular, we found 14 systems with a mass ratio less than 0.3, below the bulk of the previously known population. 
\end{sloppypar}

As some of our system have a low binding energy, it is interesting to find out if the systems are stable. Orbital evolution is possible for wide, weakly bound binaries and it might lead to their disruption as they travel through the Galaxy and encounter stars and giants molecular clouds. \cite{weinberg_dynamical_1987} worked out the calculations for the evolution and lifetimes of such wide binary systems in the solar neighbourhood; these can easily be scaled to very low-mass stars as was done by \cite{artigau_discovery_2007}. For all of our binaries, we find that the half-life is of the order of a Hubble time or larger. The probability of survival is thus high for all systems.

\begin{sloppypar}
\citet{burgasser_multiplicity_2005} have shown that the binary fraction is higher for ultra cold dwarfs that are in a wide binary system where the primary is a stellar object. Furthermore, \cite{whitworth_minimum_2006} discussed the possibility that H$_2$ dissociation might trigger a secondary fragmentation of a companion low-mass protostar if the latter is at the cooler outer parts of the circumstellar disk ($>100$~AU) and is spinning at a high enough rate. Thus, it is possible that some of our companions (and perhaps primaries) are unresolved binaries themselves. If that were the case, this could be reconciled with our common photometric distance estimates given the large associated uncertainties and could be tested observationally with either adaptive optics observations or high-precision radial-velocity monitoring of the primary.

\end{sloppypar}
To differentiate the evolutionary states of our
stars, we use the four spectral classes of metallicity defined by \cite{lepine_revised_2007} that are defined by the $\zeta_{Ti05/CaH}$ index, which is based on the  CaH2, CaH3 and TiO5 molecular bands in the optical. This index has been re-calibrated by \cite{dhital_refined_2012} and is used to differentiate dwarfs, subdwarfs, extreme subdwarfs and  ultrasubdwarfs. Using this index for all of our M dwarfs, we found that they all have $\zeta_{Ti05/CaH}$ $>$ 0.85, meaning that they all are in the dwarf metallicity class with near-solar metallicity.

\section{CONCLUSION}\label{sec:conclu}

We have discovered 14 new binary systems with companions of spectral types M6--L5 at  separations of 6\arcsec--170\arcsec\ from their primaries, corresponding to projected separations of 250--7500~AU at the distances of the systems. We also recovered nine already known binaries. Ten of our companions have a spectral type of L0 or later, two of them being comfortably in the brown dwarf regime: 2M1115+1607 (L5$\pm1$) and 2M1259+1001 (L4.5$\pm0.5$). The latter is a newly identified brown dwarf and orbits a mid-M dwarf. The most widely separated system is NLTT 687, consisting of an M3+L1 pair with a separation of 7400~AU. Other very wide systems are TYC 1725-344-1, a G5+M9 pair with a separation of 6700 AU and LSPMJ1441+1856, a M6+L1 pair with a separation of 4110 AU. Pairs consisting of a G-type star and an ultracool dwarf (e.g., HD21746, G5+M6.5) provide an opportunity to calibrate the metallicity scale of M ultracool dwarfs \cite{rojas-ayala_metal-rich_2010}. This calibration is useful because determining the metallicity of such M dwarfs would normally require parallaxes and high-resolution spectra, which are expensive data to acquire and are limited to a few bright cool stars. While our systems are not the most extremes, they can nevertheless help better define the parameter space in which wide low-mass companions can form.

We thank our referee, Eric Mamajek, for excellent suggestions that improved the quality of this paper. This research has made use of the SIMBAD database, operated at CDS, Strasbourg, France and of the VizieR catalogue access tool, CDS, Strasbourg, France. This publication makes use of data products from the Two Micron All Sky Survey, which is a joint project of the University of Massachusetts and the Infrared Processing and Analysis Center/California Institute of Technology, funded by the National Aeronautics and Space Administration and the National Science Foundation. This publication makes use of data products from the Wide-field Infrared Survey Explorer, which is a joint project of the University of California, Los Angeles, and the Jet Propulsion Laboratory/California Institute of Technology, funded by the National Aeronautics and Space Administration. This research has benefitted from the M, L, T, and Y dwarf compendium housed at \url{DwarfArchives.org}.  This publication has made use of the Very-Low-Mass Binaries Archive housed at \url{http://www.vlmbinaries.org} and maintained by Nick Siegler, Chris Gelino, and Adam Burgasser. This research has benefitted from the SpeX Prism Spectral Libraries, maintained by Adam Burgasser at \url{http://pono.ucsd.edu/~adam/browndwarfs/spexprism}.






\bibliographystyle{apj} 
{\footnotesize
\bibliography{biblio} } 

\clearpage


\end{document}